\documentclass[a3paper]{jpconf}
\usepackage{graphicx}
\usepackage{hyperref}
\usepackage{url}
\usepackage{epstopdf}
\usepackage{amsmath}
\usepackage{float}
\usepackage[pdf]{pstricks}
\begin{document}
\title{Estimation of non-femtoscopic effects in p+p and p+A collisions at RHIC energies
  using PYTHIA and HIJING generators}

\author{E Khyzhniak$^1$, N Ermakov$^1$, G Nigmatkulov$^1$}

\address{$^1$ National Research Nuclear University MEPhI (Moscow
  Engineering Physics Institute), Kashirskoe highway 31, Moscow, 115409, Russia}
\ead{evkhizhnyak@mephi.ru, noermakov@mephi.ru, ganigmatkulov@mephi.ru}

\begin{abstract}
  The spatial extents of particle emission source in high-energy collisions can be measured
  using two-particle femtoscopic correlations. In collisions with small multiplicities, such as
  proton-proton collisions, correlation functions can be distorted by non-femtoscopic effects,
  for example due to the correlations that caused by energy-momentum conservation laws, jets and mini-jets.
    To estimate these effects, a simulation of p+p collisions at $\sqrt{s}$=200 and $\sqrt{s}$=510~GeV using PYTHIA~6.4.28
  and HIJING~1.383, and p+Au collisions at $\sqrt{s_{NN}}$=200~GeV using HIJING were performed.
  Charged pion and kaon correlation functions obtained from
  the Monte Carlo generators and their comparison to the experimental data are presented.
\end{abstract}

\section{Introduction}

The spatio--temporal structures of particle emitting source in
high-energy collisions are essentially defined by the dynamics
of the collision processes~\cite{lcms_0}. The femtoscopy method
allows to measure the  spatial and temporal characteristics of
emitting region in high-energy collisions.
Such correlations arise from the quantum statistics, Coulomb and
strong final state interactions.

In small system collisions, such as p+p collisions, Bose-Einstein  
correlations can be distorted by non-femtoscopic effects, for
example, due to the correlations that caused by energy-momentum conservation laws, jets and mini-jets.

In this proceeding, estimation of jets (mini-jets) contribution to the correlation function
of pions and kaons in p+p $\sqrt{s}$=200 and $\sqrt{s}$=510~GeV, and
p+Au $\sqrt{s_{NN}}$=200~GeV collisions using PYTHIA~6.4.28~\cite{pythia} and HIJING~1.383~\cite{hijing}
Monte-Carlo generators is presented.

\section{Femtoscopy}

The method of femtoscopy is created to measure the space-time extents
of the particle emitting region at kinetic freeze-out. It is based on
measurements of identical particle quantum statistical correlations. The
femtoscopic correlations are calculated as a function of relative
momentum, expressed as $Q_{inv} = |\mathbf{p_{1}} - \mathbf{p_{2}}|$.
In order to estimate the particle emitting source parameters, one
uses the correlation function, $C(Q_{inv})$ which is constracted as: 
\begin{equation}
  C(Q_{inv}) = \frac{A(Q_{inv})}{B(Q_{inv})} ,
  \label{eq2}
\end{equation}
where $A(Q_{inv})$ is a distribution of two-particle relative
momentum that contains quantum statistical correlations, and
$B(Q_{inv})$ is the reference distribution that has all
experimental effects as the first one except for the absence of the
Bose-Einstein correlations.

One can study the dynamics of the collision evolution via measurement of the
pair transverse momentum ($k_{T} = \dfrac{|\mathbf{p_{1T}} + \mathbf{p_{2T}}|}{2}$)
dependence of the correlation function~\cite{siny}.

\section{Results and Discussions}
For estimation of non-femtoscopic effects the HIJING and PYTHIA with
Perugia~2011 (no color reconnections) and Perugia~2012 tunes were used.

\begin{figure}[H]
  \centering
  \includegraphics[width=0.55\textwidth] {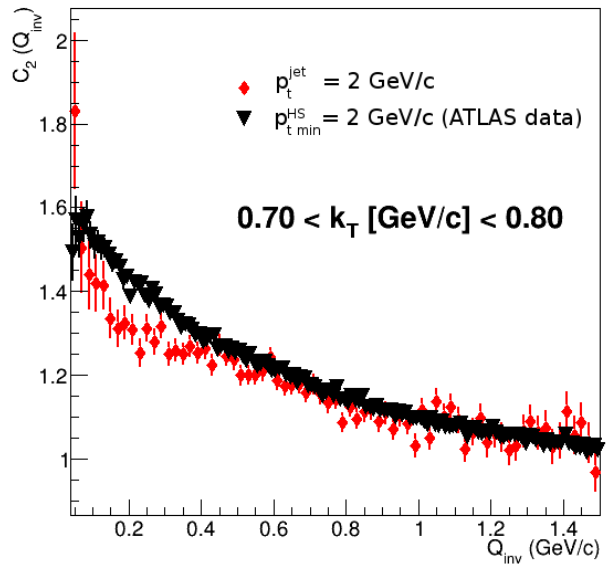}
    \caption{(Color online) Comparison of charged particle correlation functions obtained by the
    ATLAS Collaboration (black triangles) and in this analysis
    (red rombs) from HIJING with transverse momentum $0.7 < k_{T} < 0.8$~GeV/c,
    using events with a generated multiplicity $26 < N_{trk} < 36$~\cite{atlas}.}
  \label{fig1}
\end{figure}

The PYTHIA program is a standard tool for the generation of events in
high-energy collisions between elementary particles, comprising a
coherent set of physics models for the evolution from a few-body
hard-scattering process to a complex multiparticle final state.
The HIJING is a Monte Carlo event generator is designed in particular to
study jet and mini-jet production and associated particle production in
high energy p+p, p+A and A+A collisions.
The HIJING generator was running in the jet trigger mode with the
transverse momentum
of hard and semi-hard scatterings of $p_{t}^{jet}=$ 2 GeV/c.
To test the procedure, we
compared our calculations
with ones performed by the ATLAS collaboration~\cite{atlas}.
The comparison was
performed in the
multiplicity region of $26 < N_{trk} < 36$ and pair 
transverse momentum $0.7 < k_{T} < 0.8$ GeV/c.
Fig.~\ref{fig1} shows the comparison of charged particle correlation functions obtained by the
ATLAS Collaboration (black triangles) and our simulations
(red rombs). One can see, that our
calclations are consistent with published data by the ATLAS Collaboration.

\begin{figure}[H]
  \centering
  \includegraphics[width=0.8\textwidth] {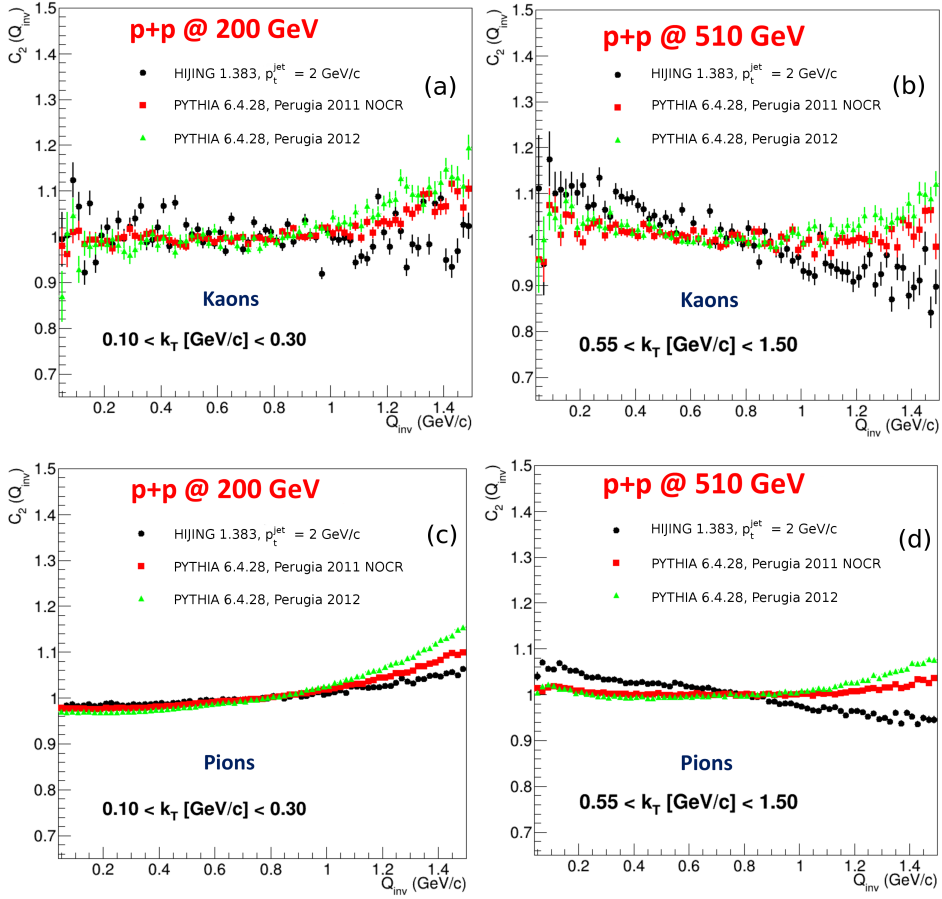}
    \caption{(Color online) Comparison of the two-kaon (a,b) and two-pion (c,d) correlation functions
    in p+p collisions at $\sqrt{s}=$200~GeV (a,c) and 
    $\sqrt{s}=$510~GeV (b,d). Black circles, red
    squares and green triangles
    represent HIJING, PYTHIA Perugia 2011 and 
    Perugia 2012 tunes, respectively.}
  \label{fig2}
\end{figure}

In Fig.~\ref{fig2} one can see the calculated two-kaon and
two-pion correlation functions at different $k_{T}$ bins
in p+p collisions at $\sqrt{s}=$200 and $\sqrt{s}=$510~GeV.

It is seen that HIJING and PYTHIA shows similar contribution of jets
(mini-jets) to the correlation functions of pions and kaons.
At low $k_{T}$ PYTHIA generator and HIJING give similar
description of non-femtoscopic correlations for both p+p
at $\sqrt{s}=$200~GeV and $\sqrt{s}=$510~GeV.
At high $k_{T}$ correlation functions obtained from 
HIJING show enhancement at low $Q_{inv}$
due to larger than PYTHIA jet (mini-jet) contribution.

\begin{figure}[H]
  \centering
  \includegraphics[width=0.55\textwidth] {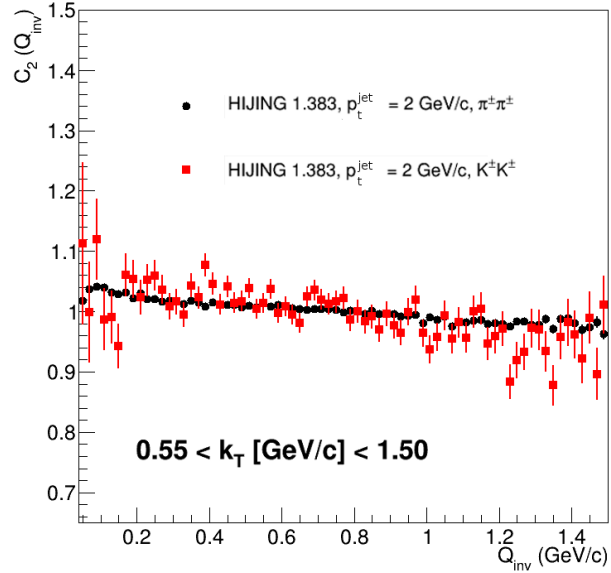}
    \caption{(Color online) Comparison of the two-kaon (black circles) and two-pion (red squares) correlation functions
    in p+Au collisions at $\sqrt{s_{NN}}=$200~GeV obtained from the HIJING generator.}
  \label{fig3}
\end{figure}

The collision system dependece of the non-femtoscopic effects and especially jets and mini-jets contribution to 
the correlation function is also very important.
Fig.~\ref{fig3} shows the two-pion and two-kaon correlation functions in p+Au collisions
at 200~GeV obtained from the HIJING generator.

One can see that at high $k_{T}$ correlation functions of charged pions obtained from HIJING show
similar trends for both p+p (Fig.~\ref{fig2}(c)) and p+Au (Fig.~\ref{fig3}) collisions.
In p+Au collisions charged pions and kaons shows similar behavior of the correlation functions.

\section{Conclusions}
The HIJING~1.383 and PYTHIA~6.4.28 is used to estimate the correlation functions
of like-sign pions and like-sign kaons in p+p collisions at $\sqrt{s}=$200~GeV
and $\sqrt{s}=$510~GeV and p+Au at $\sqrt{s_{NN}}=$200~GeV.
At low $k_{T}$ PYTHIA and HIJING generators give similar description of
non-femtoscopic correlations for both p+p $\sqrt{s}=$200~GeV and $\sqrt{s}=$510~GeV.
At high $k_{T}$ correlation functions obtained from HIJING generator shows enchacement
at low $Q_{inv}$ due to larger then PYTHIA jet (mini-jet) contribution.
In p+Au collisions charged pions and kaons show similar behavior of the correlation
functions.

\section*{Acknowledgments}
The reported study was funded by RFBR according to the 
research project No. 16-02-01119~a.
This work was partially supported by the Ministry of Science and Education of the Russian Federation, grant N 3.3380.2017/4.7, and by the National Research Nuclear University MEPhI in the framework of the Russian Academic Excellence Project (contract No. 02.a03.21.0005, 27.08.2013).
\section*{References}
\bibliographystyle{iopart-num}
\bibliography{Khyzhniak_ICPPA2017.bib}
\end{document}